\pdfoutput=1

\documentclass[11pt,a4paper]{article}
\usepackage{jinstpub}

\title{Monte Carlo characterization of PETALO, a full-body liquid xenon-based PET detector}

	\author[f]{J. Renner} 
	\author[a]{C. Romo-Luque}
	\author[c]{R.J.~Aliaga} 
	\author[b]{V.~\'Alvarez}
	\author[b]{F.~Ballester}
	\author[d]{J.M. Benlloch-Rodr\'{i}guez}
	\author[a]{J.V. Carri\'{o}n}
	\author[d]{D. Cubero}
	\author[a]{J.~D\'iaz}
	\author[b]{R.~Esteve}
	\author[b]{R. Gadea}
	\author[l]{J. Generowicz}
	\author[h]{J. Gillam} 
	\author[d]{J.L. L\'opez-G\'omez} 
	\author[a]{A.~Mart\'inez}
	\author[d, e]{F. Monrabal} 
	\author[a]{M. Querol}
	\author[g]{M. Rappaport} 
	\author[b]{J.~Rodr\'iguez}
	\author[a]{J.~Rodr\'iguez-Ponce}
	\author[i]{P. Solevi}
	\author[a]{S. Teruel-Pardo}
	\author[b]{J.F.~Toledo}
	\author[b]{R.~Torres-Curado}%
	\author[b]{V. Herrero-Bosch}
	\author[d, e]{J.J. G\'omez-Cadenas}
	\author[d, e, 1]{P. Ferrario}\note{Corresponding author.}
	\emailAdd{paola.ferrario@dipc.org}

	\affiliation[a]{Instituto de F\'isica Corpuscular (IFIC), CSIC \& Universitat de Val\`encia,\\
	Calle Catedr\'atico Jos\'e Beltr\'an, 2, 46980 Paterna, Valencia, Spain}
	\affiliation[b]{Instituto de Instrumentaci\'on para Imagen Molecular (I3M), CSIC \& Universitat Polit\`ecnica de Val\`encia, \\
	Camino de Vera s/n, Valencia, E-46022, Spain}
	\affiliation[c]{Instituto Universitario de Matem\'atica Pura y Aplicada (IUMPA), Universitat Polit\`ecnica de Val\`encia, \\
	Camino de Vera s/n, Valencia, E-46022, Spain.}
	\affiliation[d]{Donostia International Physics Center (DIPC), \\ Paseo Manuel Lardizabal 4, Donostia-San Sebasti\'an,	E-20018, Spain}
	\affiliation[e]{Basque Foundation for Science (IKERBASQUE), Bilbao, E-48013, Spain}
	\affiliation[f]{Instituto Gallego de F\'isica de Altas Energ\'ias (IGFAE), Univ.\ de Santiago de Compostela, Santiago de Compostela, Campus sur, \\
	R\'ua Xos\'e Mar\'ia Su\'arez Nu\~nez, s/n, Spain}
	\affiliation[g]{Weizmann Institute of Science, Herzl St 234, Rehovot, Israel}
	\affiliation[h]{Land Division, Defence Science and Technology Group, Fishermans Bend, Melbourne, Australia. \\ Part of the work conducted in support of this publication was while J. Gillam was at the Brain and Mind Centre, Faculty of Health Sciences, University of Sydney, NSW 2050, Australia}
	\affiliation[i]{is with Institut für Medizintechnik, Otto-von-Guericke Universität, Magdeburg, Germany (now at IBA Dosimetry GmbH, Schwarzenbruck, Germany)}
	\affiliation[j]{FULL BODY INSIGHT, S.L. Plaza Juan de Ribera 7-A, 46520, Puerto de Sagunto, Spain}

\abstract{
New detector approaches in Positron Emission Tomography imaging will play an important role in reducing costs, lowering administered radiation doses, and improving overall performance. PETALO employs liquid xenon as the active scintillating medium and UV-sensitive silicon photomultipliers for scintillation readout. The scintillation time in liquid xenon is fast enough to register time-of-flight information for each detected coincidence, and sufficient scintillation is produced with low enough fluctuations to obtain good energy resolution. The present simulation study examines a full-body-sized PETALO detector and evaluates its potential performance in PET image reconstruction.
}

\begin{document}

\maketitle

\section{Introduction}
The use of liquid xenon (LXe) as a scintillating medium presents several potential advantages in Positron Emission Tomography (PET). Liquid xenon has a high scintillation yield ($\sim$31\,000 photons per 511 keV gamma \cite{Ni:2006zp}) and a fast scintillation time. Scintillation in LXe has two components, originating from the singlet and triplet excited states, of 2.2 ns and 27 ns decay time respectively and a longer component due to recombination effects in the absence of an electric field \cite{Kubota_1979}. Moreover, LXe is a continuous medium with uniform response, therefore distortions due to border effects are expected to be reduced and the reconstruction of the position of gamma interactions improved. Being a continuous detector, a full 3D reconstruction of the gamma interactions is possible, which improves the precision of the determination of the line of response (LOR) along which the gammas are emitted. In addition to an excellent time resolution, this can lead to a boost in the quality of reconstructed images. PETALO is a novel detector concept for PET imaging based on LXe and a silicon photomultiplier (SiPM) read-out, which is currently exploring these potential advantages \cite{Gomez-Cadenas:2016mkq, Gomez-Cadenas:2017bfq, Renner:2020ayj}. 


Full-body PET detectors present a considerable number of advantages over smaller scanners \cite{Cherry_2018}. Their larger size increases the geometrical coverage and therefore the overall sensitivity, which could be realized in shorter image acquisition times, lower administered tracer doses, and improved image signal-to-noise ratio. Imaging the entire body at once avoids having to make multiple scans targeting different regions of the body. These benefits combine to yield further advantages, as less time between the tracer injection and the end of the acquisition procedure permits more efficient use of the tracer decay time, shorter acquisitions mean less potential for error due to patient movement during the scan, and an overall lower dose rate leads to fewer random coincidences. The recently-constructed EXPLORER detector is the first full-body PET/CT scanner. It employs LYSO crystals over a 194 cm axial field-of-view and has produced scans of an entire human body \cite{Badawi_2019}.  Additional gains in overall PET sensitivity have been shown to result from the use of Time-of-Flight (TOF) in image reconstruction \cite{Conti_2019, Westerwoudt_2014}.  
We envision a similar detector constructed with a LXe active medium, which has also the advantage of being extremely competitive in terms of cost, thanks to the scalability of the concept of a continuous scintillator. It has already been suggested \cite{Gomez-Cadenas:2016mkq, Gomez-Cadenas:2017bfq} that an excellent Coincidence Time Resolution (CTR) down to $\sim$ 100 ps can be achieved in a small set-up made of two boxes of $2.4\times 2.4\times 5$ cm$^3$, with highly reflective walls, which provides an optimal light collection, and SiPMs sensitive to the VUV xenon scintillation light. 

Here we present a simulation study of a full-body PET based on the PETALO concept, from the resolution achievable in the determination of the high energy gamma interaction in xenon (in energy, position and time) to the image reconstruction of a point with an iterative Maximum Likelihood Expectation Maximization algorithm.


\section{The PETALO concept}
\begin{figure}[htbp]
	\centering
	\centerline{\includegraphics[width= 0.9\textwidth]{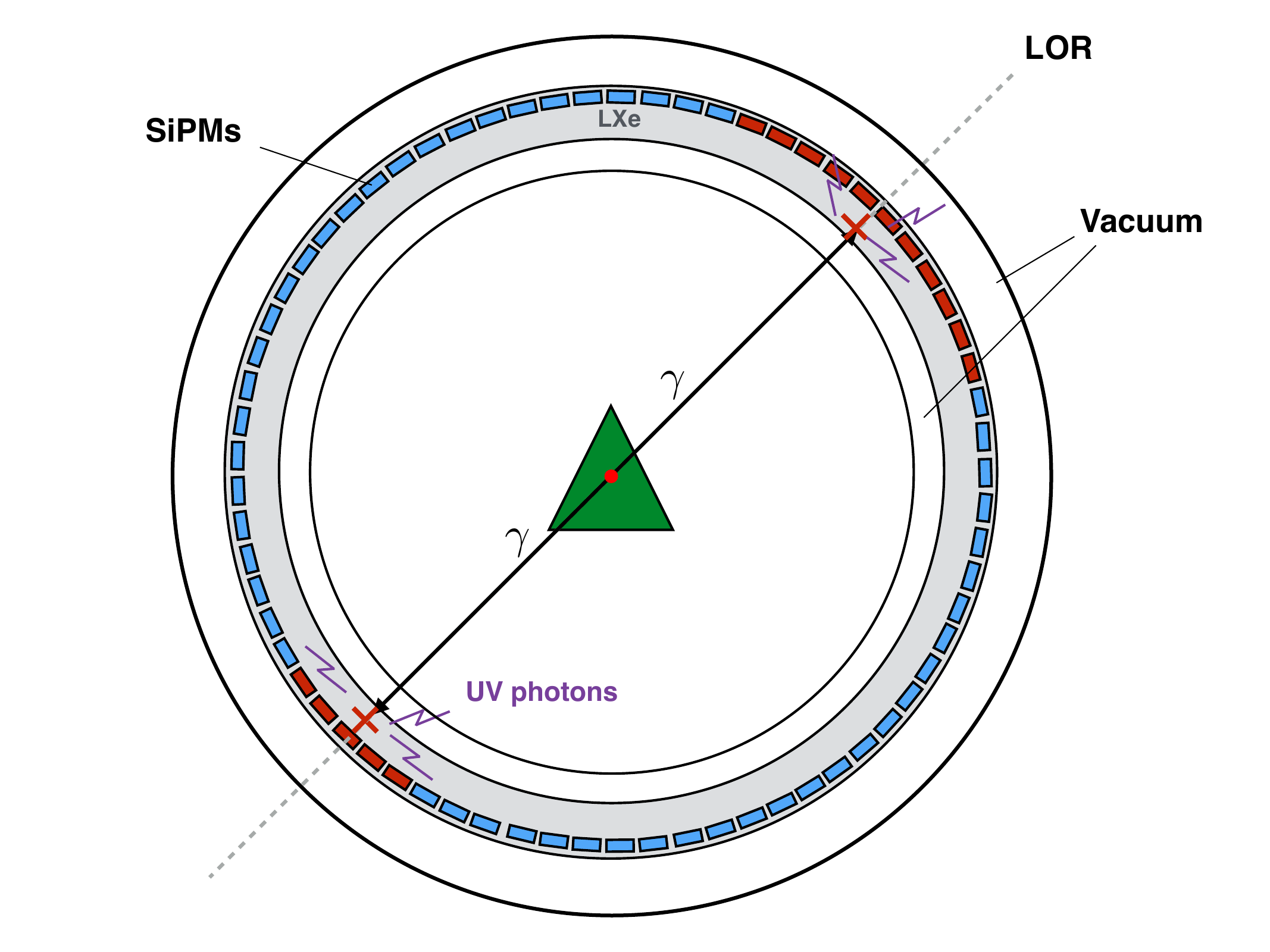}}
	\caption{Geometry of a PETALO detector. A vacuum-isolated cylindrical ring of LXe is instrumented on the external radially inward-facing surface with arrays of SiPMs which observe the UV scintillation produced by the $e^+e^-$ annihilation radiation emitted in the active region. Notice that the figure is not at scale.}
	\label{fig.petalogeom}
\end{figure}

The PETALO concept consists of a ring-like structure filled with LXe and instrumented internally with SiPMs (see Fig.~\ref{fig.petalogeom}). 
SiPMs cover the external inner-facing wall with normals in the radial direction. Almost collinear 511 keV gamma rays produced by $e^+e^-$ annihilation in the observed volume interact in the LXe, producing UV scintillation which is detected by the SiPMs. The resulting pattern of light can be used to determine the interaction locations of each gamma ray and thereby the line of response along which they were emitted, and the difference in arrival times of the scintillation gives the TOF difference which can be used to localize the emission along the LOR. 

LXe is contained in one single, continuous volume, inside a cryostat, which keeps the temperature around 161 K, where xenon liquefies at atmospheric pressure. Xenon is constantly recirculated in gas phase through a purification system, to eliminate impurities that could quench the scintillation light such as water and oxygen, and liquefied again upon entering the detector, passing through a heat exchanger to maximize the cooling efficiency. 

The sensors are large-area (typically $6\times6$ mm$^2$) VUV-sensitive SiPMs. The performance of such devices has improved greatly over the last few years, while their cost has been steadily reduced.  As an example, the MPCC S15779 model from Hamamatsu has a photon detection efficiency (PDE) of 30\% at 178 nm, a factor 3 higher than the first models which appeared in the market less than five years ago. 

The front-end and read-out \cite{IEEE2018talk} are based on TOFPET2 ASICs \cite{petsys}, which provide early digitization of sensor outputs and a very fast timestamp, thanks to the usage of two different thresholds, a very low one to get an accurate timestamp and a second one to validate the signal.



\section{Simulation and reconstruction}
\label{simreco}

A complete simulation including the emission of almost collinear 511 keV gamma rays, their interactions in a full-body PETALO geometry, and the propagation and detection of UV photons emitted in the LXe active region is generated using the \textsc{Geant4} software toolkit \cite{Geant}. The scanner is 195 cm long, has an inner radius of 38 cm and and a thickness of LXe of 3 cm. This is a value close to the attenuation length of 511-keV gammas (at this depth around 60$\%$ of photons will be stopped), to maintain a good efficiency, and not too large, not to introduce large errors in the reconstruction. The scintillation properties of LXe have been modelled in detail, including the scintillation spectrum, its decay constants, the refraction index as a function of the photon wavelength, and the Rayleigh scattering. A summary of the most relevant properties used in the simulation can be found in Table \ref{t:simprop}. 
\begin{table}[h!]
\centering
\begin{tabular}{@{}l|l@{}}
Density & 2.953 g/cm$^3$ \\
Attenuation length for 511 keV gammas & 3.7 cm \\
Scintillation yield for 511 keV gamma & $\sim 30\ 000$ photons \\
Peak scintillation wavelength & 178 nm \\
Peak refractive index & 1.69 \\
Rayleigh scattering length &  36.4 cm \\
Effective decay constant $\tau_1$ &  2 ns\\
Effective decay constant $\tau_2$ & 43.5 ns \\
 \end{tabular}
\caption{LXe relevant properties used in the simulations.}\label{t:simprop}
\end{table}
The scintillation decay time is modelled according to the measurements in Ref.~\cite{Hogenbirk_2018}, as the sum of two exponentials of constants 2 and 43.5 ns, with relative intensity $3\%$ and $97\%$, respectively. The longer constant includes the effect of recombination, which appears in the absence of an electric field.

The way SiPMs are modelled has a strong impact on the amount and the fluctuation of detected light, which affects also the time reconstruction. In this simulation we use sensors with an active area of $6\times6$ mm$^2$ protected by a window made of a specific type of quartz, which transmits 90\% of VUV photons in vacuum, and has a refractive index of 1.6 to LXe scintillation light, following the properties of custom-made arrays of SiPMs of the series S15779 of Hamamatsu, which are currently being used in the first PETALO prototype. 

Finally, the amount of collected light on the SiPMs and its distribution has also an important impact on the event reconstruction. For this study we have assumed that all internal surfaces that are not instrumented absorb the scintillation photons perfectly. Although this means that less light is collected by the SiPMs overall, it allows one to identify better the 3D position of the gamma interactions, because the reflected light does not contain any information on its origin. The SiPM coverage of the external surface of the cylinder is around 73 $\%$, while the overage coverage is 37 $\%$.

For a single event:

\begin{enumerate}
	\item[1.] A vertex is generated in the desired volume.
	\item[2.] Two 511 keV gamma rays are launched from the vertex along a randomly oriented line. The deviation angle from collinearity is simulated according to a gaussian distribution with sigma equal to 0.23 deg.
	\item[3.] The gamma rays are propagated in the PETALO geometry and any interactions in the active region are recorded as energy depositions (or ``hits''). UV scintillation photons are also generated according to the deposited energy in the LXe and are individually propagated through the geometry. 
	\item[4.] The number of photons striking each SiPM is recorded and a photodetection efficiency of 30\% assumed, to store the number of detected photons in time.
	\item[5.] The SiPM shaping is simulated, convoluting the content of each time bin with a shaping function, consisting of the sum of two exponential functions, and a threshold is applied to simulate the timestamp assignment of the ASIC.
\end{enumerate}

\subsection{Estimation of energy and position resolution}
\label{reco_performance}

\begin{figure}[htbp]
	\begin{center}
	\includegraphics[width= 0.49\textwidth]{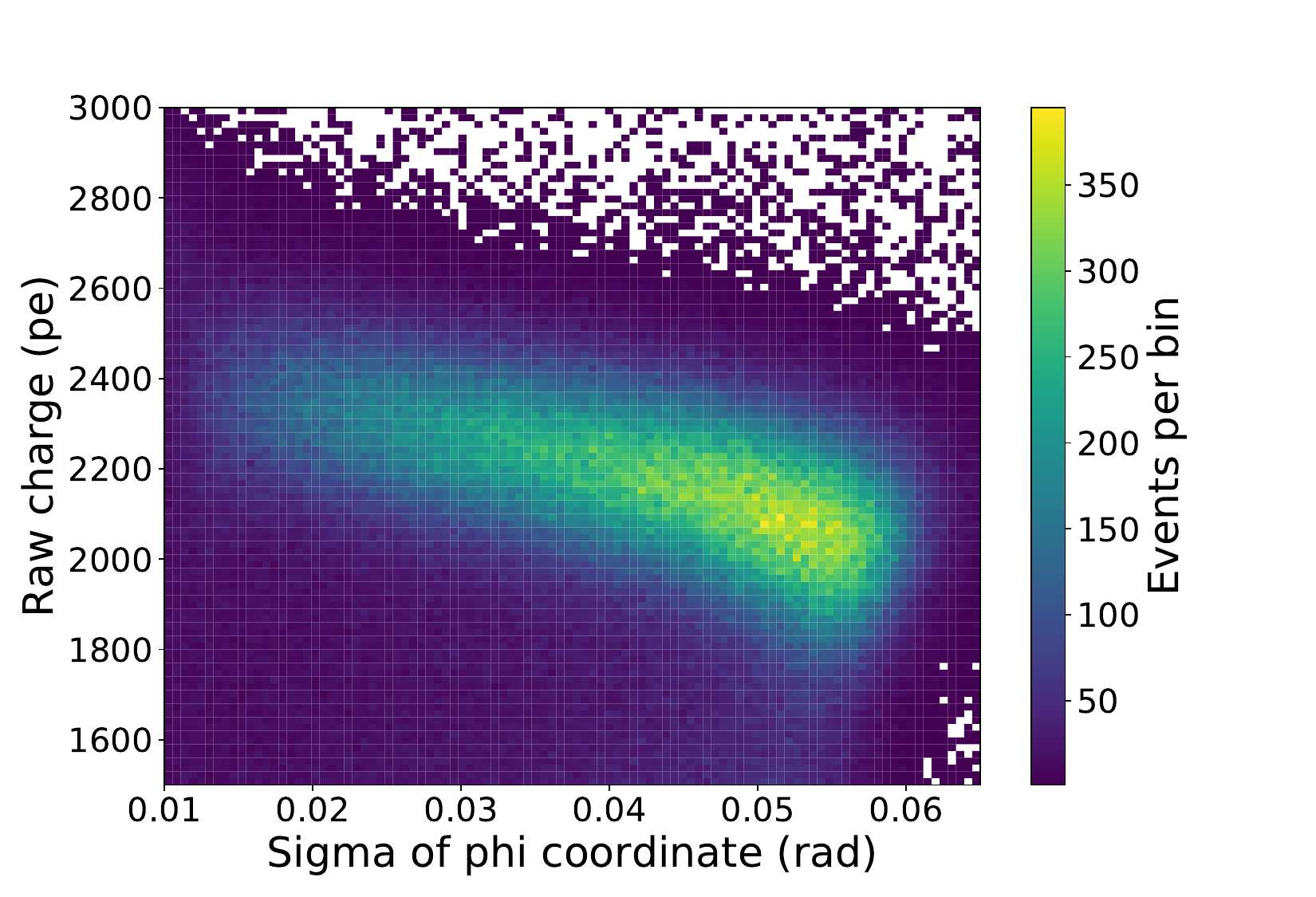}
	\includegraphics[width= 0.49\textwidth]{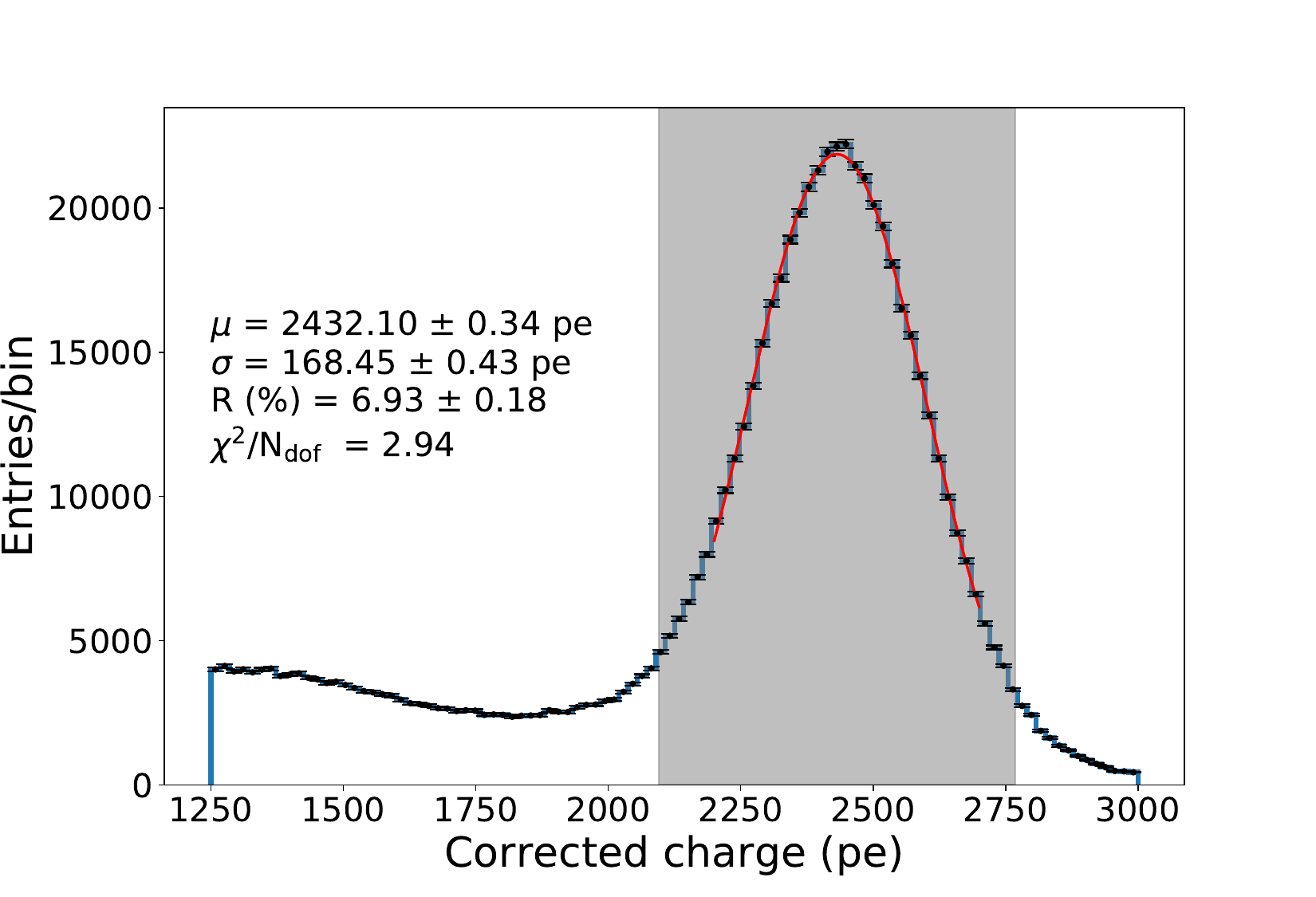}
	\caption{Left: dependence of the raw charge on the charge spread in the angular coordinate. Right: total charge detected by the SiPMs of each half of the detector. The visible peak is formed by gammas depositing all their energy in the detector, either through one single photoelectric interaction or one or more Compton scatters followed by a photoelectric interaction. The region around the photo-peak is shown, and a fit is applied to extract the resolution. The grey band shows the range of the coincidence filter subsequently applied.}
	\label{fig.raw_energy}
	\end{center}
\end{figure}

To characterize the detector in terms of energy, position and time resolution, simulations from a point-like source in the center of the scanner have been studied.

In the absence of charge collection, energy resolution in liquid xenon is dominated by the non-proportionality of the scintillation light with the energy deposited and the fluctuations due to recombination in the absence of an electric field. The first effect implies that events with the same total deposited energy present different collected charge if the event is a single photoelectric or a Compton plus a photoelectric interaction. The estimated amount of these two intrinsic fluctuations has been measured to be $6\%$ $\sigma$  \cite{Ni:2006zp} and it has been added to the charge distribution provided by the \textsc{Geant4} simulation, fluctuating the detected charge by that amount, since this effect is not included in the toolkit.

Also, the amount of collected charge on the SiPMs depends on the radial position of the interaction point: the closer it is to the sensors, the larger the detected charge. 
At the same time, photons emitted farther from the SiPMs produce a larger spread of the light cone, which can be quantified as the standard deviation of the angular coordinate:

\begin{equation}\label{eqn.phirms}
\phi_{\mathrm{std}} = \sqrt{\frac{1}{Q_{\phi}}\sum_{q_{i} > q_{t,\phi}}(\phi_{i} - \overline{\phi})^2q_{i}} \,
\end{equation}
where $q_{i}$  is the charge detected by the \textit{i-th} SiPM, $Q_{\phi}$ is the total charge, $\phi_i$ is the coordinate of the \textit{i-th} SiPM and $\overline{\phi}$ is the average $\phi$ coordinate.
Fig.~\ref{fig.raw_energy}-\textit{left}  shows the correlation between the amount of detected charge and the spread in the angular coordinate, which can be used to correct the charge interpolating a quadratic polynomial function.


Events in which both gamma rays have interacted in the active volume are identified based on the SiPM information. The SiPM sensor responses are divided into two halves, with a plane perpendicular to the line connecting the position of the maximum-charge SiPM and the center of the system, and passing through the center. If both SiPM sets have a total charge within $\pm 2 \sigma$ around the photoelectric peak (between 2095 and 2769 photoelectrons), the event is selected as a coincidence. In Fig.~\ref{fig.raw_energy}-\textit{right}  the total charge recorded by each subset of SiPMs is shown, once corrected for the dependence on the spread of light on the SiPMs, as well as the energy selection, illustrated by the grey band. The energy resolution is dominated by the intrinsic fluctuation due to the non-proportionality of scintillation light in LXe, and results in $\sim6.9\%$ $\sigma$ or 16.2$\%$ FWHM. 
%
\begin{figure}[htbp]
	\begin{center}
	\includegraphics[width= 0.9\textwidth]{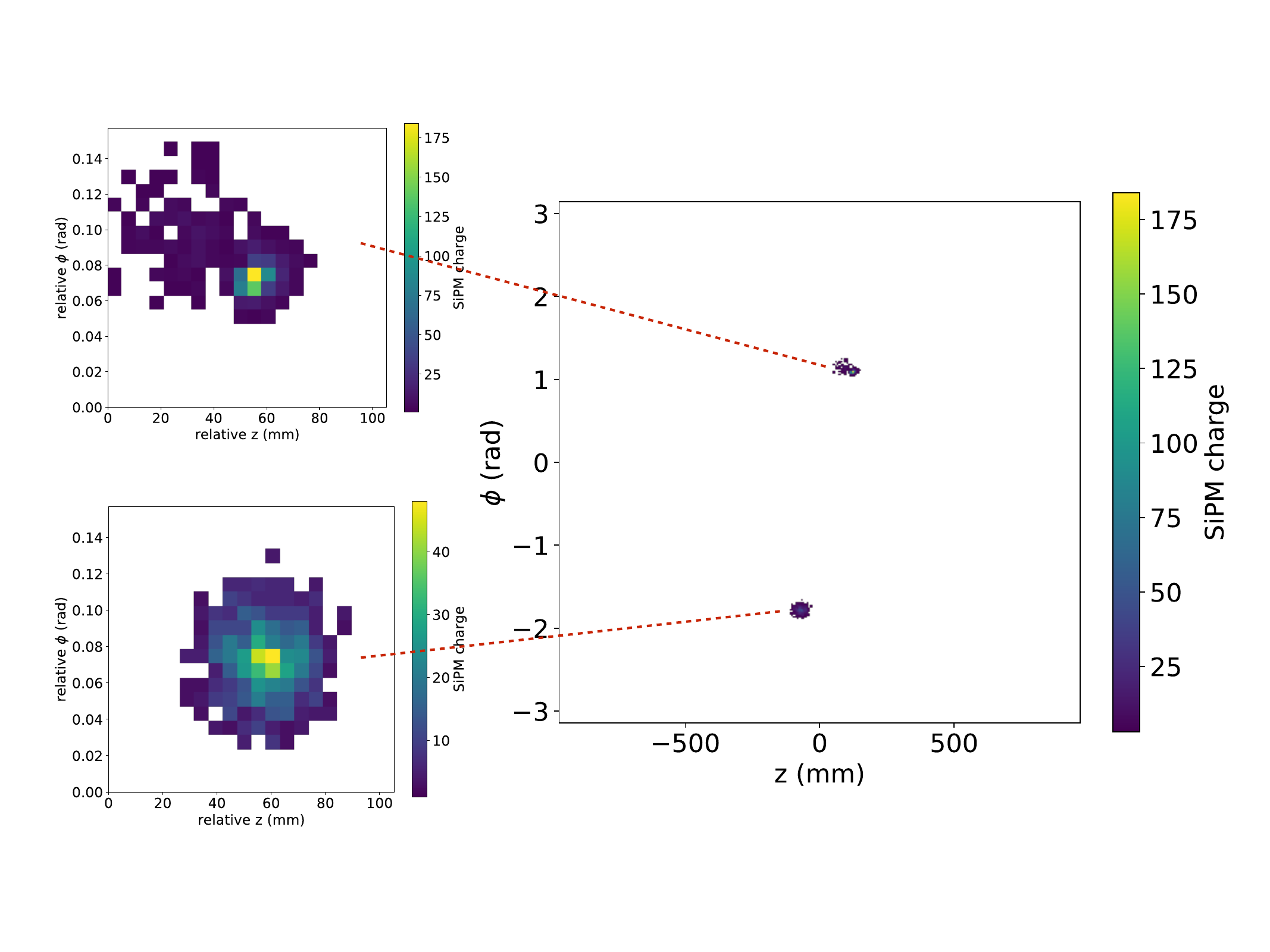}
	\caption{Simulated SiPM responses (in detected photons) for an event containing two coincident 511 keV gamma rays. The sensor responses produced by each gamma are clearly separated, and the sensor locations and detected charges in the two separate regions (shown enlarged at left) are used to reconstruct $(r,\phi,z,t)$ coordinates for each gamma ray.}
	\label{fig.reco}
	\end{center}
\end{figure}
The fraction of coincidence events over the total number of simulated events is 0.081, giving a sensitivity of 81 counts/second/kBq for a point in the centre of the field of view. Since the typical activity for radiotracers used in PET scanners can be of several MBq, any background such as muons or environmental radioactivity is negligible.

\begin{figure}[htbp]
	\begin{center}
	\includegraphics[width= 0.9\textwidth]{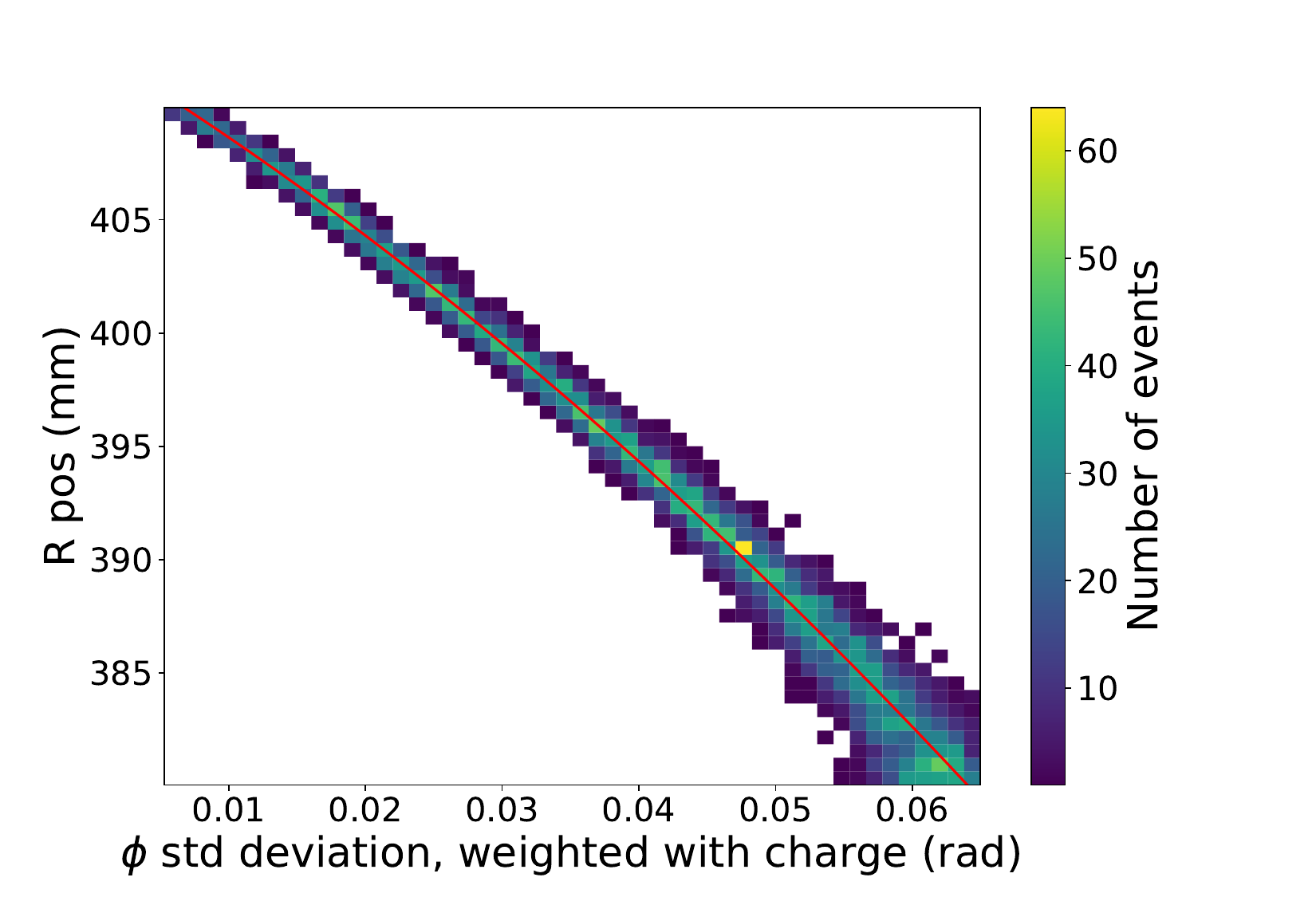}
	\caption{True radial position of gamma interactions for pure photoelectric events, as a function of the spread of the $\phi$ coordinate of the SiPMs detecting charge above a certain threshold (4 pe).}
	\label{fig.rdep}
	\end{center}
\end{figure}

In Fig.~\ref{fig.reco} an example of the SiPM response to a coincidence event is shown, where the two gamma interactions are clearly visible. For each set of SiPM responses, the coordinates $(r,\phi,z)$ of the individual gamma rays are computed. Individual charge thresholds can be set for the reconstruction of each coordinate, with the aim of cleaning up the charge pattern on the sensors, eliminating fluctuations at the border of the light cone. These thresholds have been varied to find the optimal values, which give the best resolution.

The $r$ coordinate is calculated using the radial dependence of the charge spread on the photosensors. This dependence is extracted using the Monte Carlo true information of the interaction of each gamma in the xenon, using photoelectric interactions only. In Fig.~\ref{fig.rdep} the true radial position of the gamma interaction is shown as a function of the standard deviation of the $\phi$ coordinate of the SiPMs that have detected charge above a threshold of 4 pe, weighted with the charge (see Eq.~\ref{eqn.phirms}).

 Therefore, the measurement of the charge on the SiPMs allows one to calculate the radial position. Naturally, this method reconstructs pure photoelectric events better, or, more broadly, events where all the energy of the gamma is deposited in a limited area (within a few mm), which we define as \textit{photoelectric-like}, while interactions with one or more Compton scatters in which the energy is deposited over a wider area, suffer from a larger reconstruction error.  It is worth to notice that \textit{photoelectric-like} coincidences are not only pure photoelectric events, but also Compton events in which the gammas do not travel a significant distance after scattering. On the other hand, when two or more interactions at a distance larger than few mm occur in the detector, the precision of the reconstruction worsens, because the pattern of light on the SiPMs is the result of the sum of light coming from different points, and the first photoelectron to be detected can originate in a second or third interaction, thus providing no information on the position of the first interaction, which is the relevant one for image reconstruction. These effects introduce fluctuations in the precision of the reconstructed position, spoiling the resolution. 

\begin{figure*}[htbp]
\begin{center}
	\includegraphics[width= 0.49\textwidth]{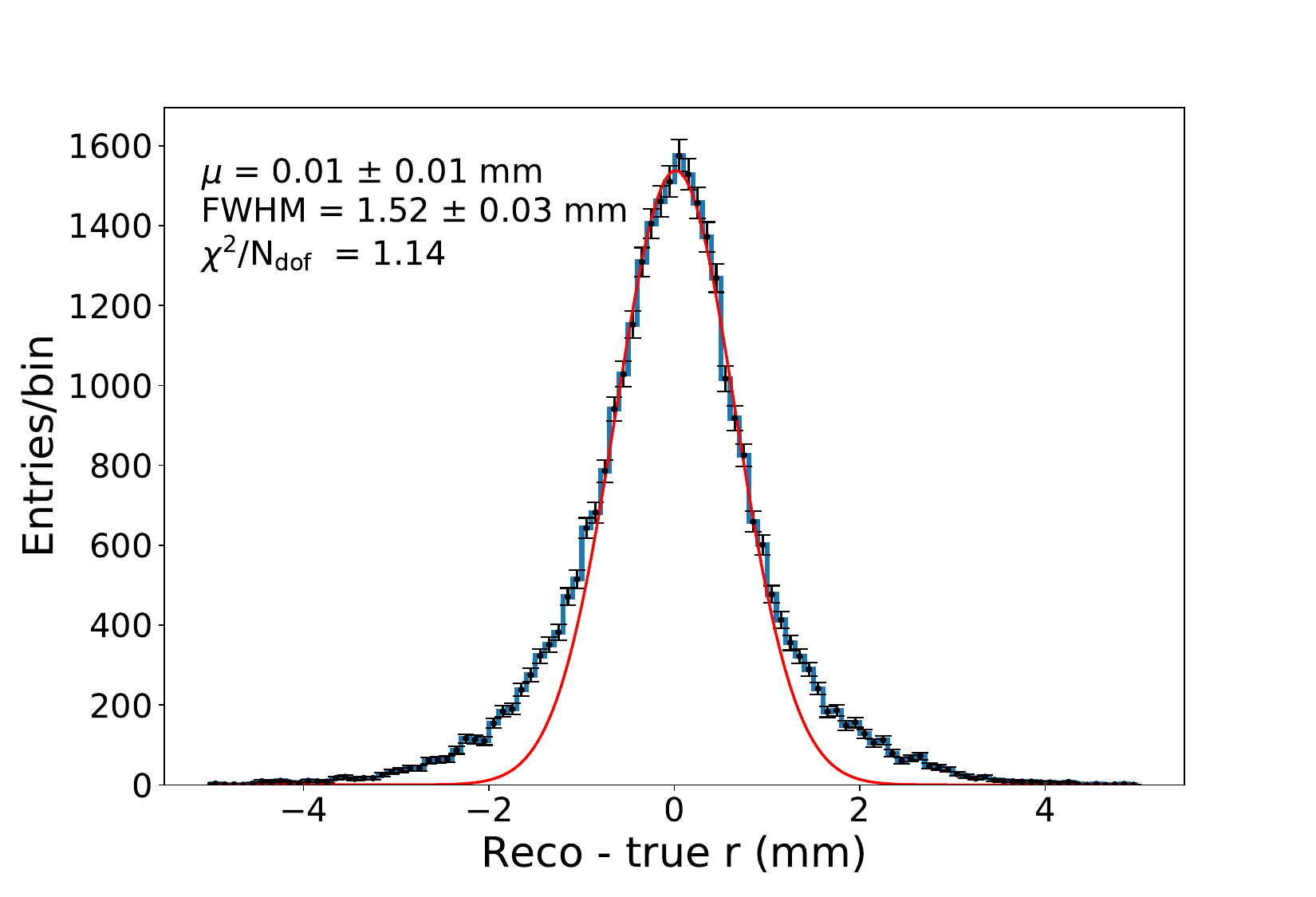}
	\includegraphics[width= 0.49\textwidth]{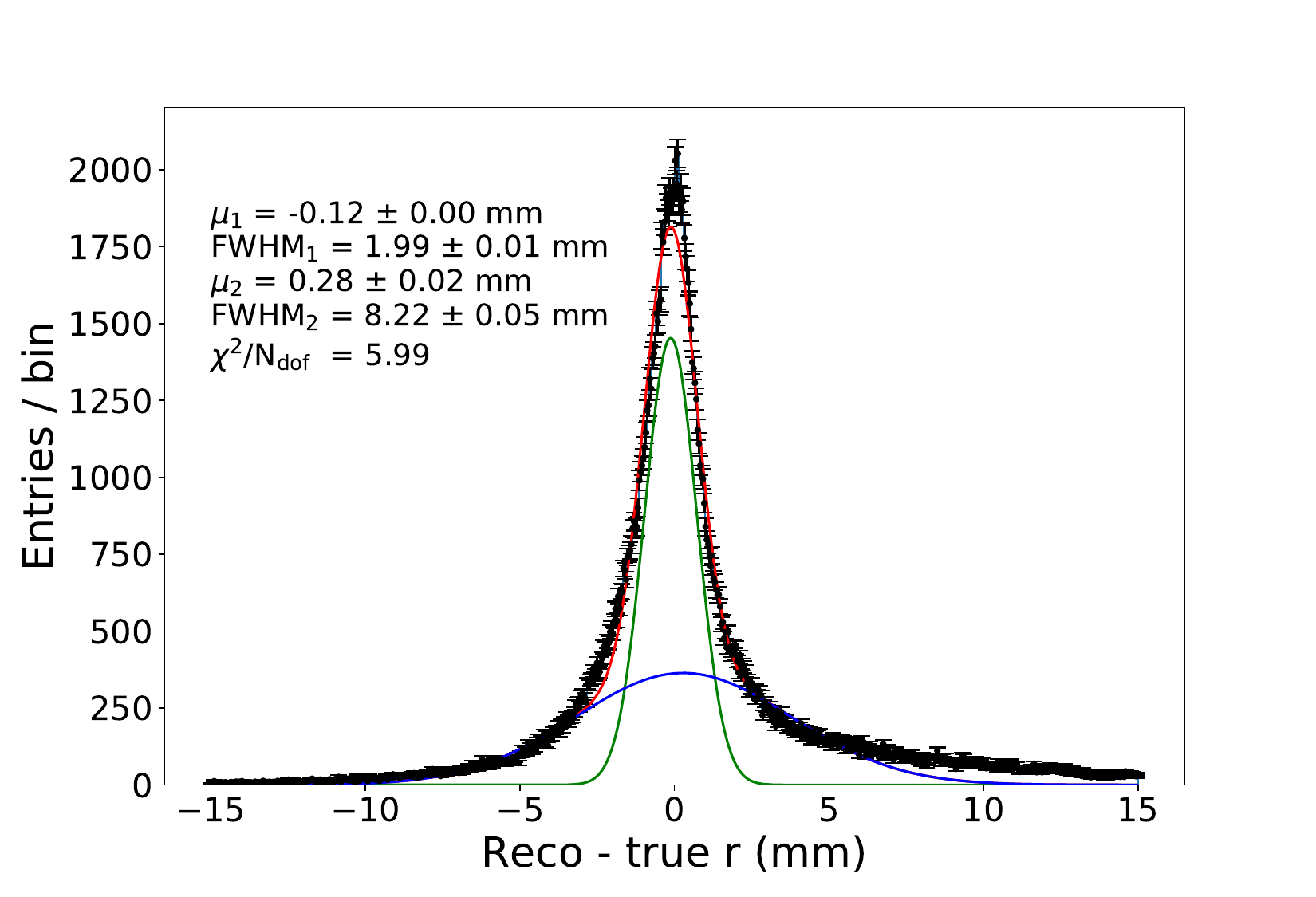}
	\caption{Difference between the reconstructed and true radial position for \textit{photoelectric-like} events (left panel) and for the whole sample (right panel). Notice that the two images have different X scale.}
	\label{fig.rres}
	\end{center}
\end{figure*}

Fig.~\ref{fig.rres} shows the error in the reconstruction of the radial coordinate for \textit{photoelectric-like} events, where all the deposited energy hits lie within 1 mm from each other (left panel) and for the whole sample (right panel). In the first case the radial error is around 1.5 mm FWHM, with the appearance of small tails, while the whole sample shows a distribution with larger tails, which can be modelled as the sum of two gaussian distributions. The narrower distribution presents a FWHM resolution of around 2 mm, while the broader one degrades to around 8 mm.

\begin{figure*}[htbp]
\begin{center}
	\includegraphics[width= 0.49\textwidth]{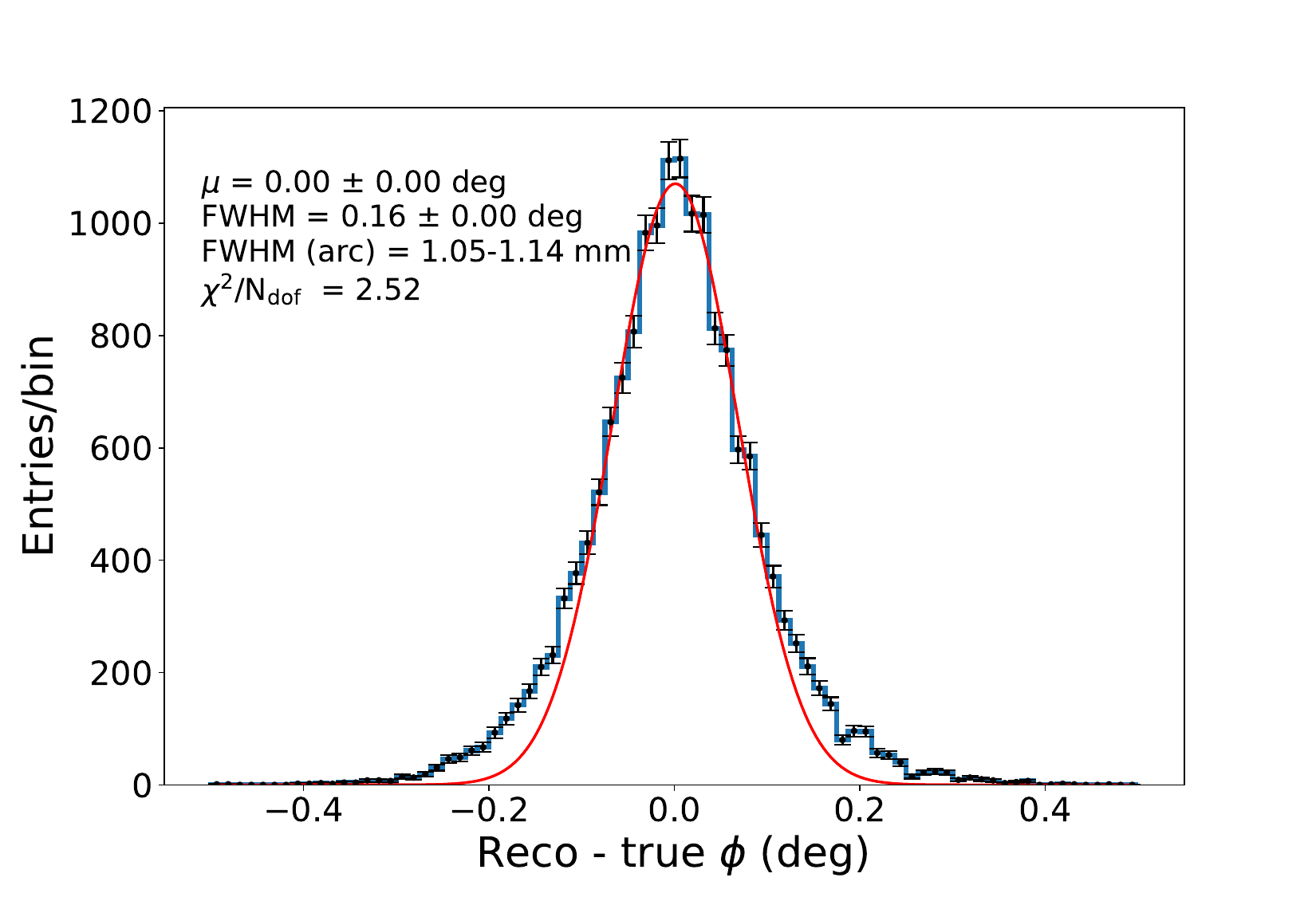}
	\includegraphics[width= 0.49\textwidth]{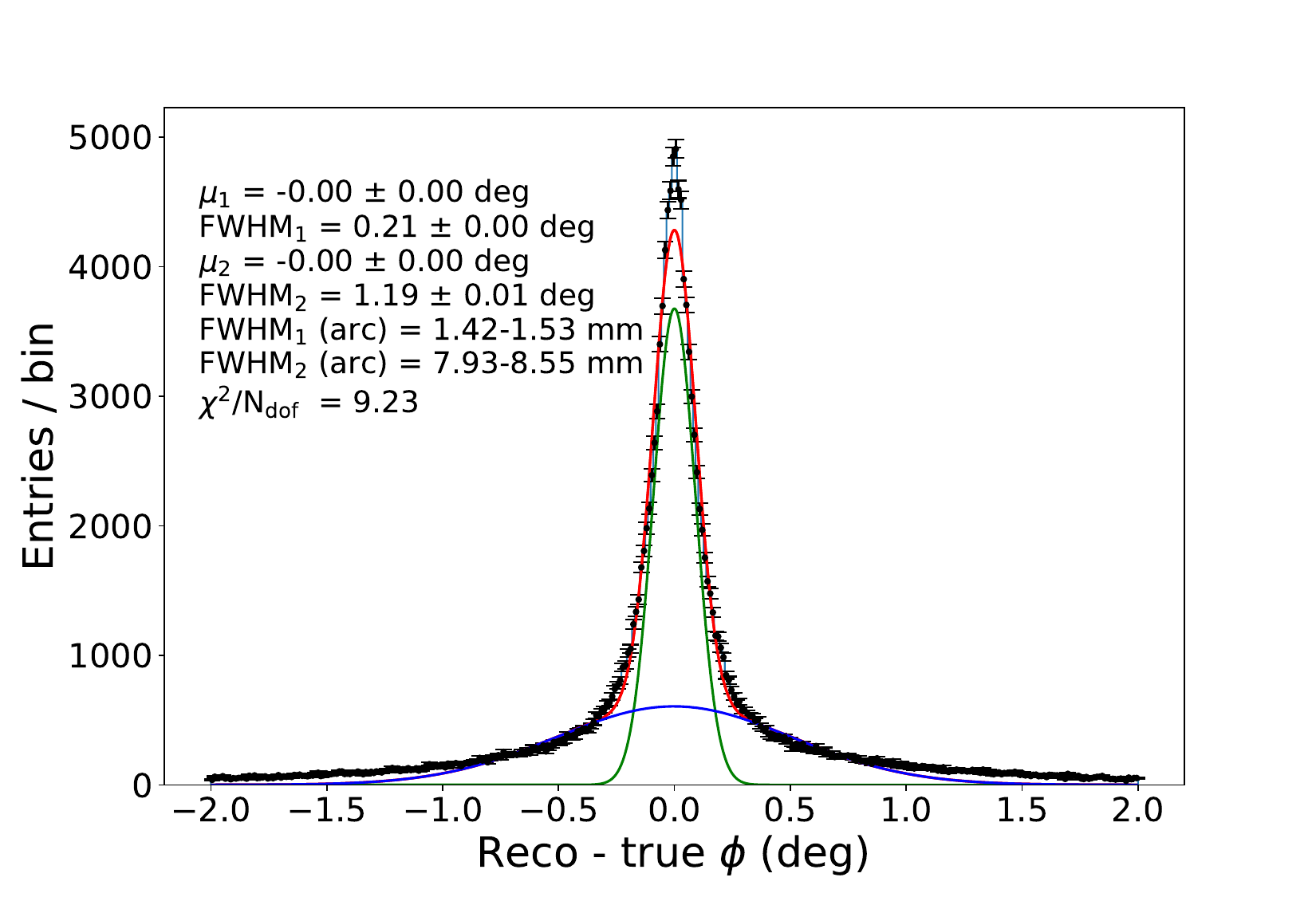}
	\caption{Difference between the reconstructed and $\phi$ coordinate for \textit{photoelectric-like} events (left panel) and for the whole sample (right panel). Notice that the two images have different X scale.}
	\label{fig.phires}
	\end{center}
\end{figure*}

\begin{figure*}[htbp]
\begin{center}
	\includegraphics[width= 0.49\textwidth]{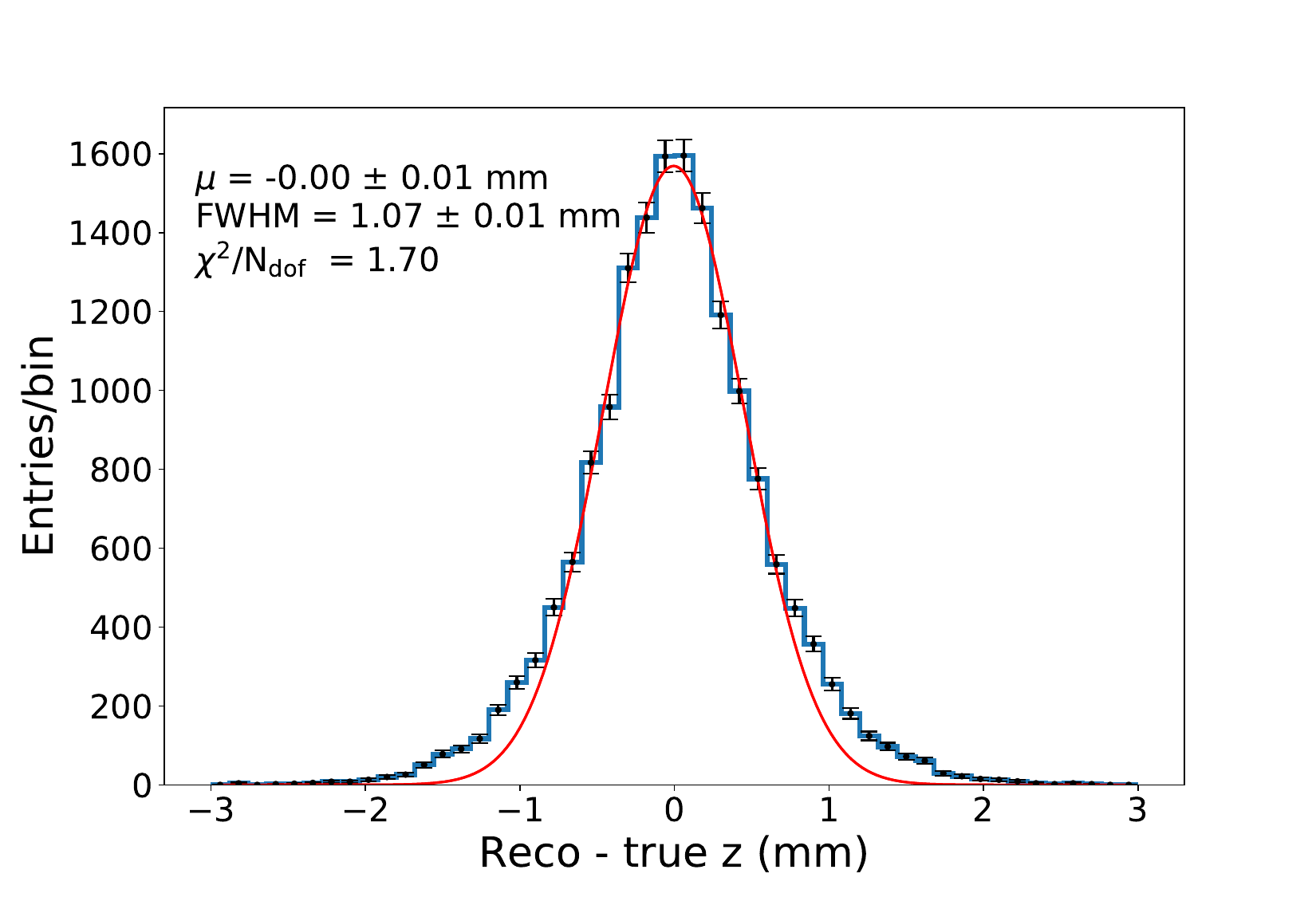}
	\includegraphics[width= 0.49\textwidth]{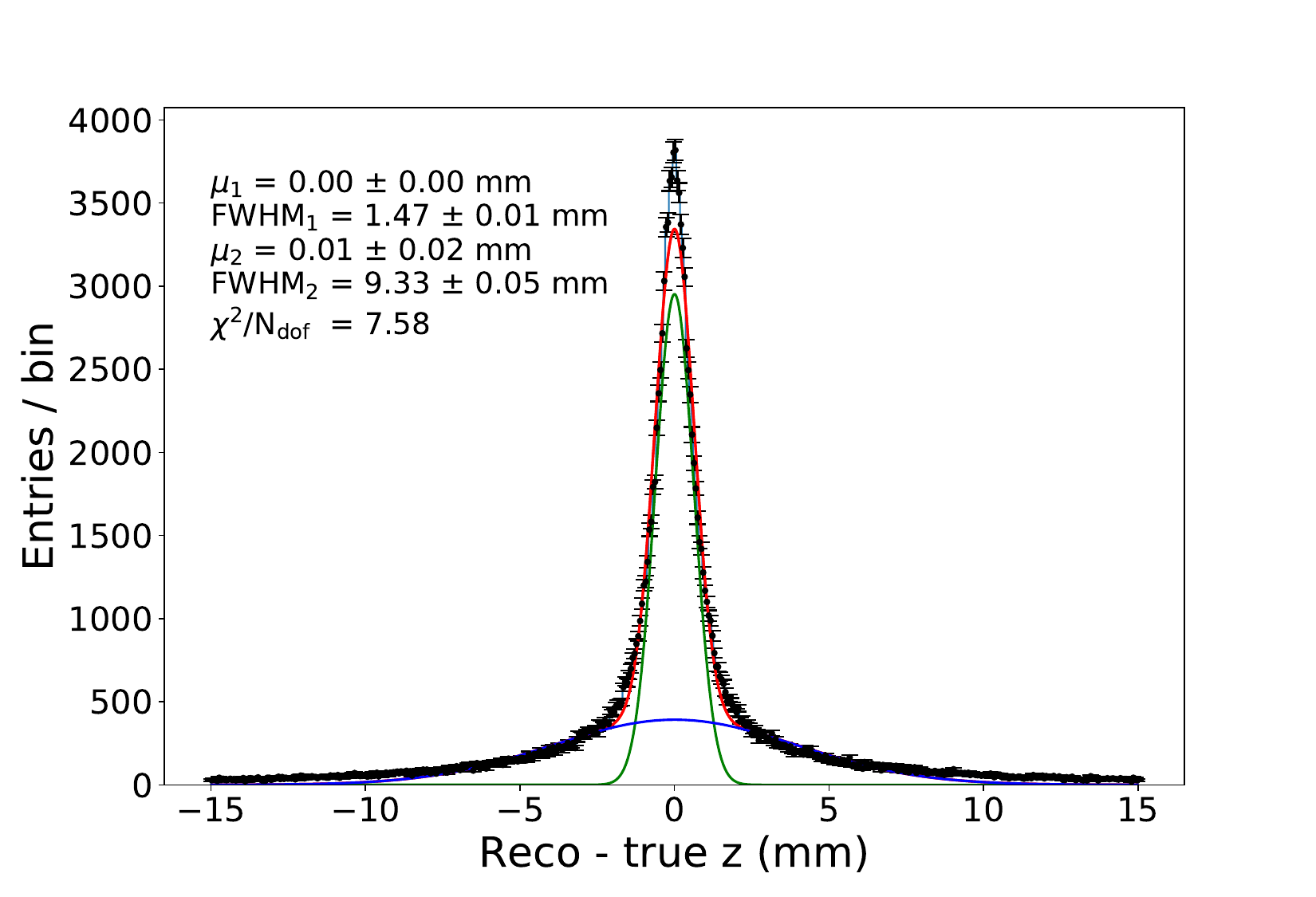}
	\caption{Difference between the reconstructed and $z$ coordinate for \textit{photoelectric-like} events (left panel) and for the whole sample (right panel). Notice that the two images have different X scale.}
	\label{fig.zres}
	\end{center}
\end{figure*}

The $z$ and $\phi$ coordinates can be calculated as

\begin{equation}\label{eqn.zphi}
z = \frac{1}{Q_{z}}\sum_{q_{i} > q_{t,z}}z_{i}q_{i},\,\,\,\, \phi = \frac{1}{Q_{\phi}}\sum_{q_{i} > q_{t,\phi}}\phi_{i}q_{i},
\end{equation}

\noindent where $z_{i}$, $\phi_{i}$, and $q_{i}$ are the $z$ coordinate, $\phi$ coordinate, and measured charge of sensor $i$.  $q_{t,z}$ and $q_{t,\phi}$ are charge thresholds, such that in each case, only sensors with $q_{i}$ greater than a threshold are used in the computation, and $Q_{z}$ and $Q_{\phi}$ are the total SiPM charges used in the calculation of $z$ and $\phi$, defined as

\begin{equation}
Q_{z} = \sum_{q_{i} > q_{t,z}}q_{i},\,\,\,\, Q_{\phi} = \sum_{q_{i} > q_{t,\phi}}q_{i}.
\end{equation}

As shown in Fig.~\ref{fig.phires} and Fig.~\ref{fig.zres} the $\phi$ and $z$ error distributions also present large tails, corresponding to Compton events. However, for \textit{photoelectric-like} events, the spatial resolution is around 1 mm for both coordinates. The charge thresholds $q_{t,z}$ and $q_{t,\phi}$ that optimize the reconstruction are found to be 4 pe for both.

\subsection{Time of Flight}
\label{tof_performance}

 \begin{figure*}[htbp]
 \begin{center}
	\includegraphics[width= 0.49\textwidth]{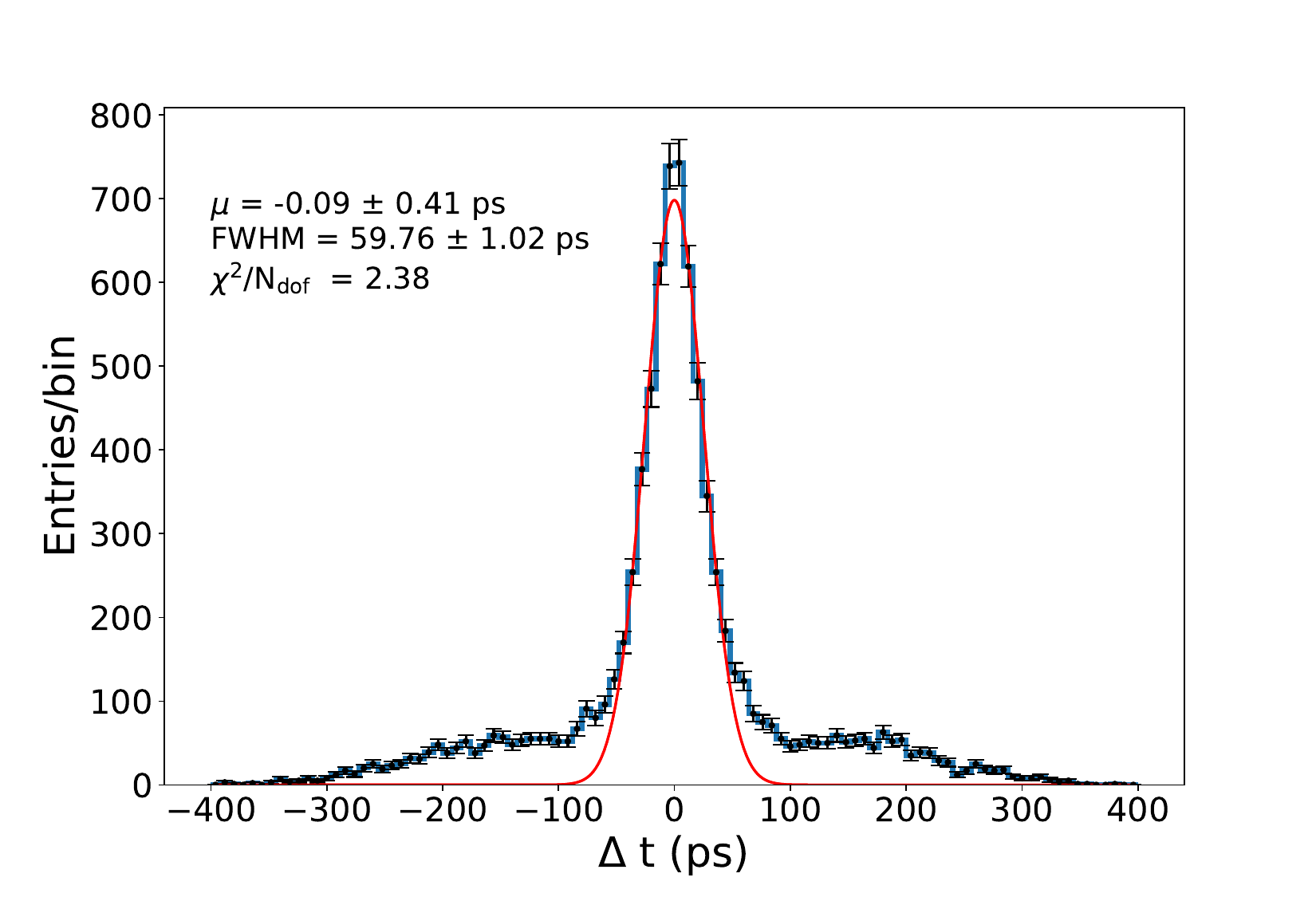}
	\includegraphics[width= 0.49\textwidth]{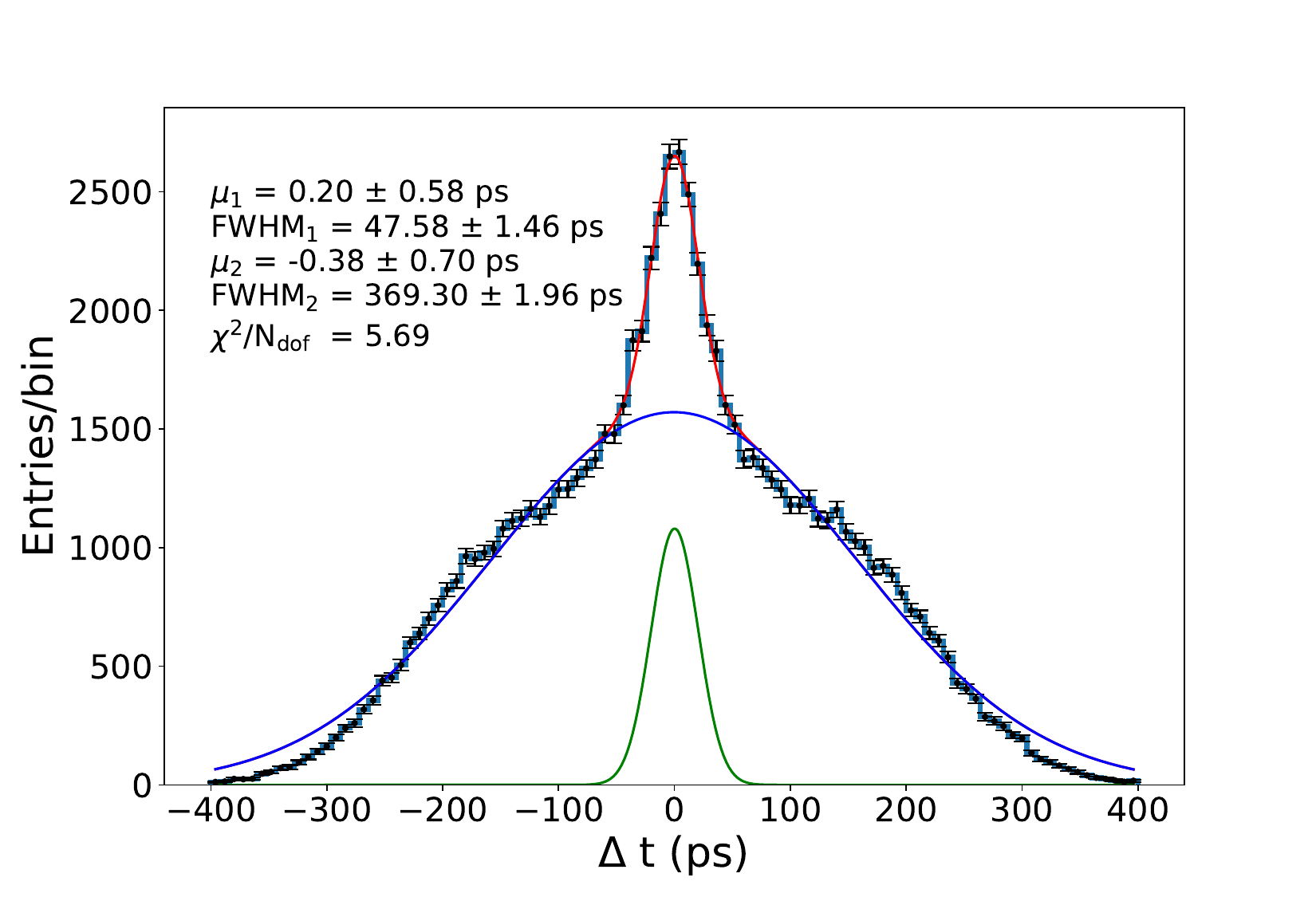}
	\caption{Coincidence Time Resolution for \textit{photoelectric-like} events (left panel) and the whole sample (right panel).}
	\label{fig.ctr}
	\end{center}
\end{figure*}

The coincidence time resolution is usually defined as the standard deviation or full-width-half-maximum of the $t_1^\mathrm{det} - t_2^\mathrm{det} $ distribution, where $t_{1(2)}^\mathrm{det} $ is the timestamp of the first detected photon for each gamma interaction. Since the gamma emission point is the center of the geometrical system, this distribution has central value equal to zero and its fluctuation is due to the position of the gamma interaction inside a crystal, the fluctuation of the emission time of the photons and the time photons take to reach a SiPM and get detected. While this is true for PET scanners with pixelated crystals, which don't provide information on the actual point where the gamma interacts inside the crystal, in a continuous, LXe-based PET $t^\mathrm{det} _{1(2)}$ can be corrected using the information on the 3D position. In the ideal case where both the time and the position of the gamma interactions were known with no error, the point of emission of the gammas would be identified with perfect resolution. Therefore, we define the CTR as the FWHM of the difference between two quantities: the difference between the measured arrival time of the first photons on the SiPMs (corrected for the time spent by optical photons to arrive to the photosensors) and the difference between the true time of the interaction of the gammas in xenon:

\begin{eqnarray}
{\rm CTR} &= & \Delta^{{\rm meas}} - \Delta^{{\rm true}} \\ 
& = & t_1^\mathrm{det}  - t_2^\mathrm{det}  - (t^{\textrm{opt}}_1 - t^{\textrm{opt}}_2) - (t^{{\rm true}}_1 - t^{{\rm true}}_2)\, ,
\end{eqnarray}
where $t^{\textrm{opt}}$ is the time the first scintillation photon takes to go from the emission point to the SiPM where it is detected and $t^{{\rm true}}$ is the time between the emission of the 511-keV gamma and its interaction in the xenon. Since the speed of optical photons in LXe is not constant, because the refraction index depends on the wavelength, but there is no information available on the wavelength of the detected photons, the value corresponding to the wavelength peak (1.69) is used to calculate $t^{\textrm{opt}}$, to optimize the correction. 


As mentioned in Sec.~\ref{simreco}, a timestamp is assigned to each SiPM, in the following way: the shaping of the SiPM is simulated convoluting each photoelectron with a function equal to the sum of two exponentials, with characteristic constants 100 and 15000 ps. Subsequently, a threshold is applied which returns the time when the SiPM response exceeds the threshold. A value of 0.25 pe has been chosen for the present study, which is the lowest threshold that TOFPET2 ASICs can reach for time measurement \cite{petsys}. The time of the SiPM waveform corresponding to this threshold is then chosen as a proxy for the time of the interaction.

As it happens with the spatial resolution, the attainable time resolution is better for \textit{photoelectric-like} events than for multi-site events. In the left panel of Fig.~\ref{fig.ctr} it can be seen that a CTR of around 60 ps FWHM is obtained for the former. In the whole sample, shown in the right panel, two distributions can be identified, which correspond to \textit{photoelectric-like} events and the rest of events. A fit to two gaussian functions can be performed, to extract the CTR for both of them, obtaining $\sim$50 ps FWHM for \textit{photoelectric-like} events and $\sim$370 ps FWHM for the rest of the sample, very good results, which show the potential of liquid xenon in TOF measurements.

It is clear from the above results that the fluctuations introduced by the sensors and the electronics can be the dominant effect in the time resolution. The main contribution of the SiPMs is given by the specific microcell where the photon is detected within a SiPM, which introduces a delay which depends on the position of the cells in the array. Currently available SiPMs show an average smearing of a $\sigma$ of $\sim$40 ps \cite{sipm_jit}. On the other hand, the electronics chain introduces a further fluctuation in time, which can be estimated to be around 30 ps $\sigma$ \cite{FundamentalLimits}.

 \begin{figure}[htbp]
 \begin{center}
	\includegraphics[width= 0.9\textwidth]{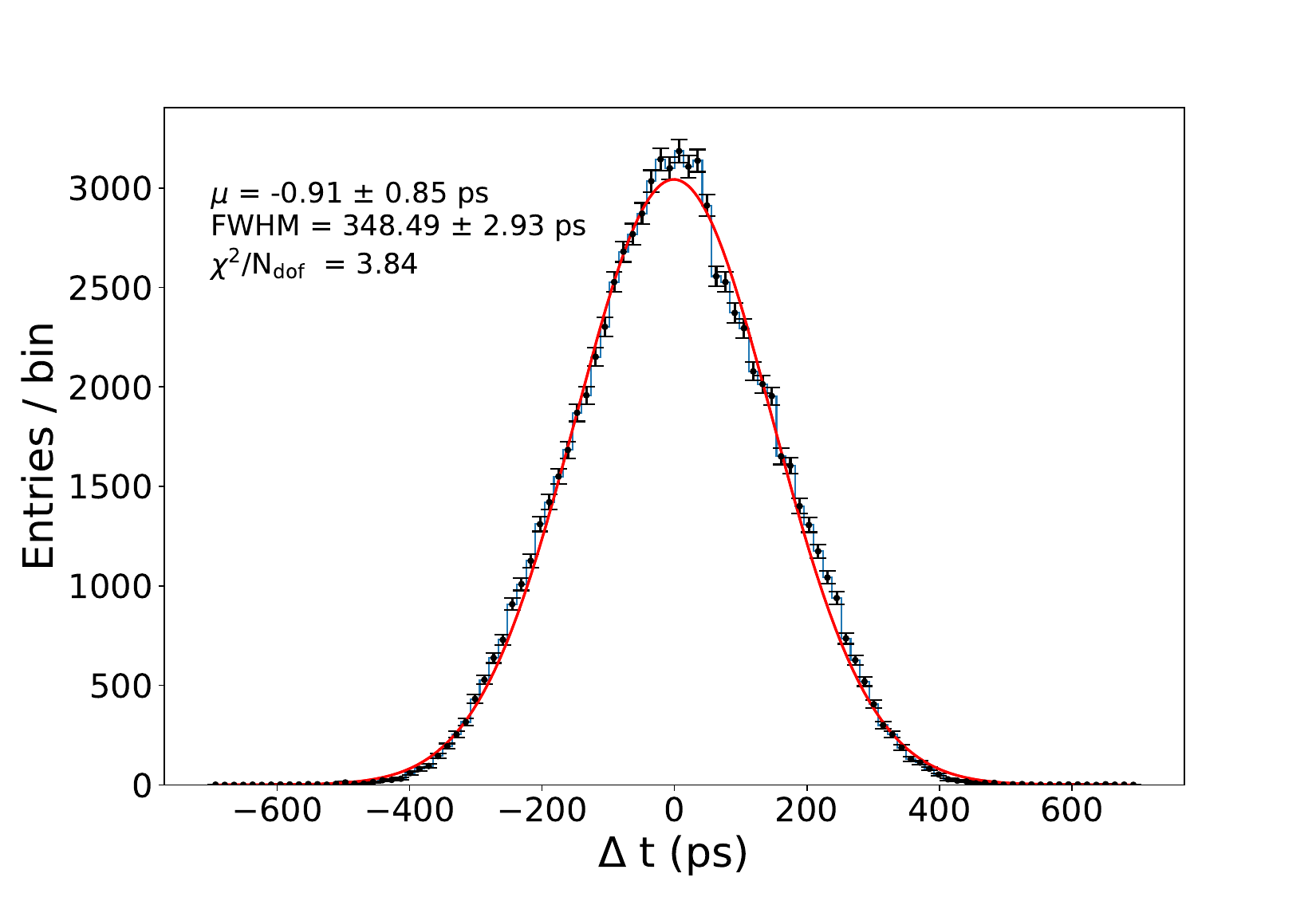}
	\caption{Coincidence Time Resolution for all the coincidences, taking into account the time fluctuations introduced by SiPMs and electronics.}
	\label{fig.ctr_jit}
	\end{center}
\end{figure}

We have studied how the CTR is affected by these two contributions, introducing a gaussian smearing in the simulation, for each one of them. The resulting CTR for the whole ensemble of events is  around 350 ps, as shown in Fig.~\ref{fig.ctr_jit}. It is interesting to notice that the difference between the two gaussian distributions, which can be appreciated in Fig.~\ref{fig.ctr}\textit{--right} (\textit{photoelectric-like} or full ensemble of events) is blurred out by the addition of the electronics and sensor fluctuations.

\subsection{Discussion}
The performance of  a full-body PET detector based on LXe and SiPM readout is excellent in the reconstruction of events where the energy of the gamma is deposited in a point-like region of 1 mm. A position resolution around 1 mm is found for the three coordinates, as well as a CTR of 50 ps, assuming a read-out with no instrumental time smearing. Increasing the threshold for the definition of \textit{photoelectric-like} events leads to more efficiency, at the expenses of only a mild worsening of the resolution. For instance, using a limit of 3 mm for the area considered as a point, 18$\%$ of the coincidences are retained, compared to 10$\%$ of 1 mm, while the resolution in the radial coordinate stays practically the same. 
One could in principle try to clean the sample filtering out Compton events based on the pattern of light detected by the photosensors. However, often no clear difference is visible in the light pattern for photoelectric and Compton events, especially when the multiple interactions happen not so far from each other or on the same radial direction. However, instead of trying to distinguish photoelectric from Compton events,  one can try to distinguish events with most energy deposited in a limited region, which means separating
 events with poorly reconstructed positions from events where the position is reconstructed well, based on the pattern of detected charge on the SiPMs.  This way an excellent performance could be achieved, at the expenses of only a reduced loss of efficiency. The possibility of identifying good versus bad reconstructed events have been studied recently using Deep Neural Networks \cite{Renner:2020ayj} and work is in progress to apply this kind of algorithms to image reconstruction.

 Time resolution is also affected by fluctuations due to the SiPM response and the electronics. With the current state-of-the-art technology, these effects dominate the resolution and the resulting CTR is around 350 ps FWHM, an extremely competitive result. As a comparison, the measured CTR for the \textsc{uExplorer} first total body PET is 505 ps \cite{review_fb}.

\section{One point resolution}
\label{point_res}

The spatial resolution of an image achievable with PETALO has been estimated reconstructing the image of a point slightly displaced from the centre of the field of view by 1 cm in the $y$ direction (0, 1 cm, 0). For the reconstruction we employ a 3D listmode maximum likelihood expectation maximization (MLEM) reconstruction algorithm with a gaussian TOF kernel \cite{tofpet3d}. This permits reconstruction of the entire 3D volume of interest beginning from the set of interaction coordinate pairs $(x_1,y_1,z_1,t_1)$, $(x_2,y_2,z_2,t_2)$. The algorithm iteratively reconstructs the image $\lambda^{(k)}_{j}$, where $j$ runs over all reconstructed 3D voxels and $k$ is the iteration number, as (see, for example, \cite{IAEA_handbook})

\begin{equation}
\lambda^{(k+1)}_j = \frac{\lambda^{(k)}_j}{\sum_{i}A_{ij}}\times \sum_{i_{m}}\Biggl(\frac{A_{i_{m}j}}{\sum_{j}A_{i_{m}j}\lambda^{(k)}_{j}}\Biggr).
\end{equation}

\noindent Here $i$ runs over all possible LORs and $i_{m}$ over all measured LORs, $A_{ij}$ is the system matrix describing the probability of emission from voxel $j$ given an observation of LOR $i$.

\begin{figure}[!htbp]
 \begin{center}
 	\includegraphics[width= 0.46\textwidth]{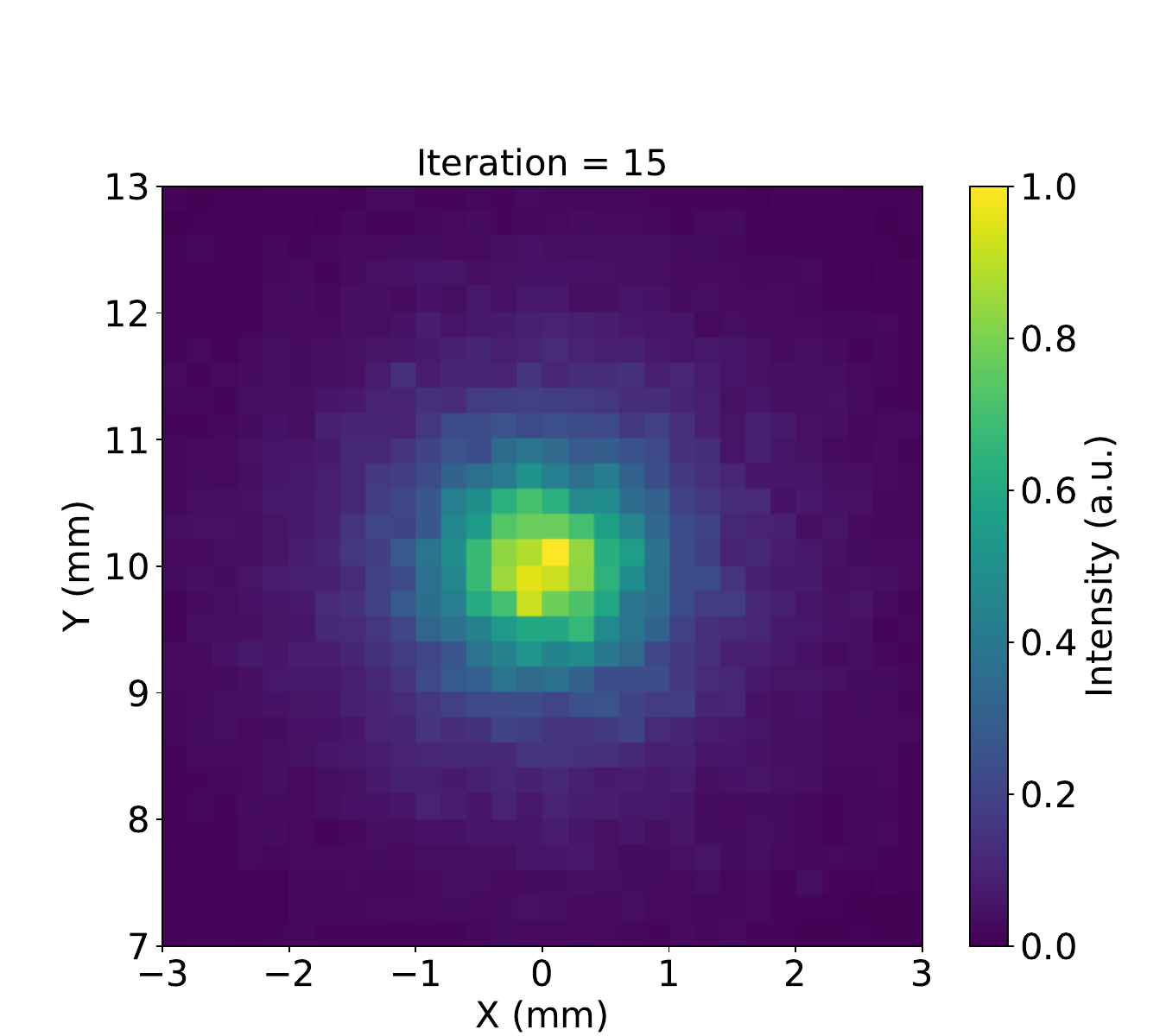}
	\includegraphics[width= 0.52\textwidth]{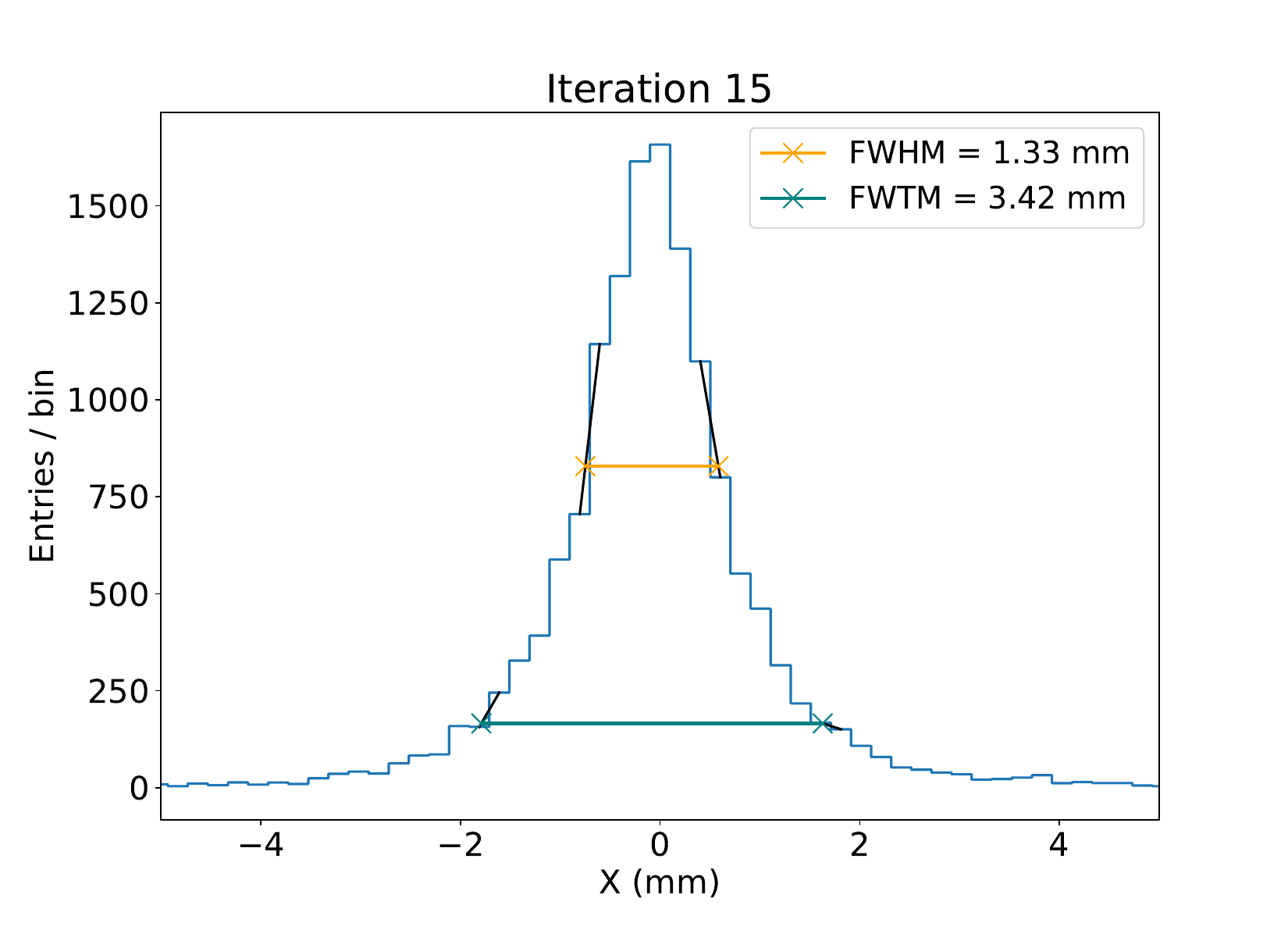}
	\includegraphics[width= 0.52\textwidth]{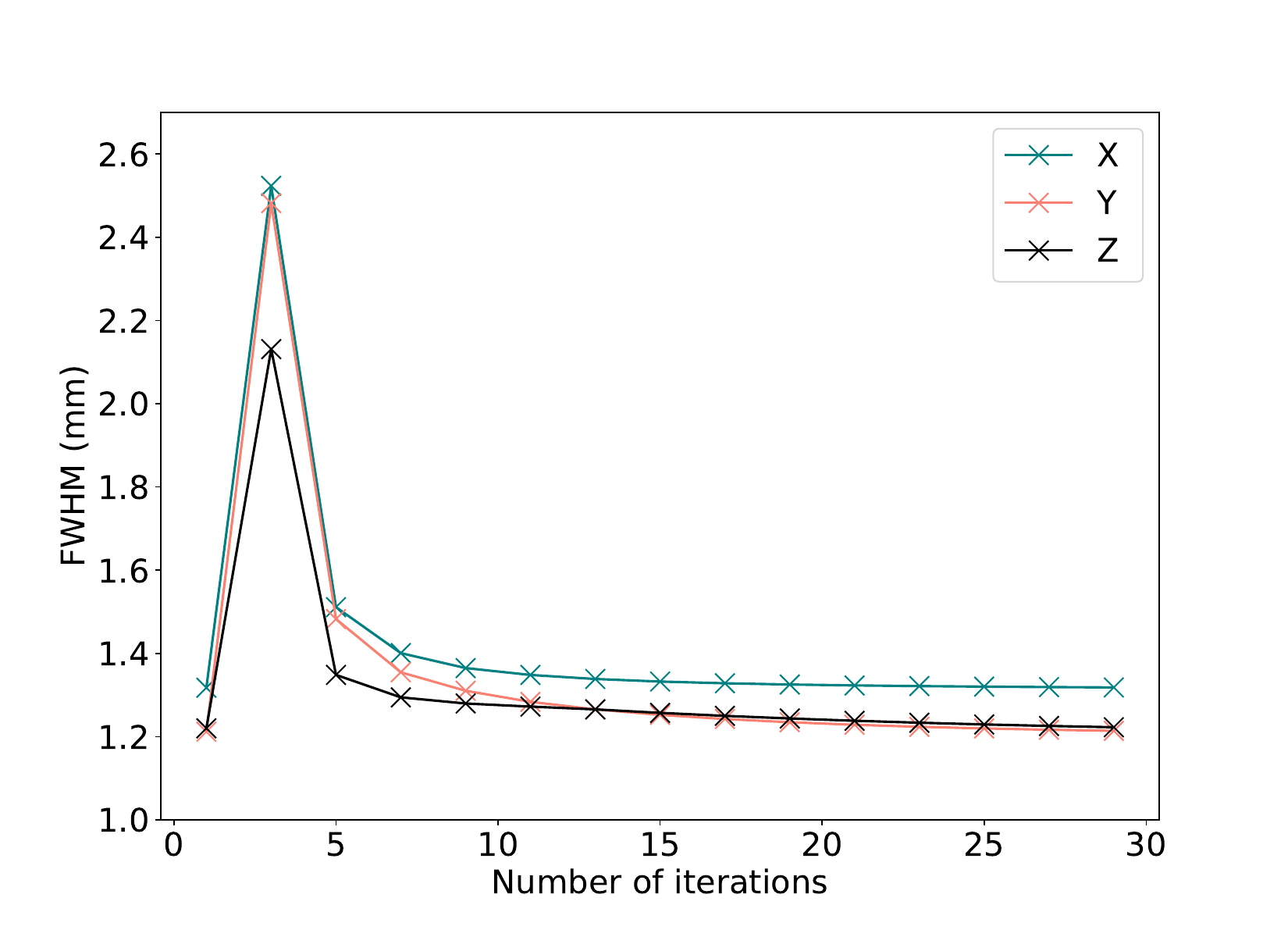}
	\caption{Top: Image (left) and profile (right) in the $x$ coordinate of the reconstructed image of a point in (0, 1 cm, 0) from the center of the field of view, after 15 iterations. Bottom: resolution of the reconstructed point as a function of the number of iterations used in the MLEM algorithm for the three coordinates.}
	\label{fig.pointres3cm}
	\end{center}
\end{figure}

In Fig.~\ref{fig.pointres3cm}\textit{--top-left} the reconstructed image of the point is shown in the $xy$ plane that passes through the peak of the distribution. The FWHM of the point source image is determined for three orthogonal directions ($x$, $y$, $z$) along profiles through the image volume, passing through the peak of the distribution. Each FWHM is calculated by linear interpolation between adjacent pixels at half the maximum value of the image intensity (Fig.~\ref{fig.pointres3cm}\textit{--top-right}.). We found that the resolution converges to around 1.3 mm FWHM for the $x$ and $y$ coordinates, after a few iterations and 1.22 mm FWHM for the $z$ coordinate, as shown in Fig.~\ref{fig.pointres3cm}\textit{--bottom}. To understand the main contributions to the blurring of the reconstructed image, we have also reconstructed the image using simulations with completely collinear gammas; in that case the point resolution is 0.7 mm FWHM, which means that non collinearity is the dominant contribution.

\section{Conclusions}
\label{sec.conclusions}

In this work, we have studied for the first time the potential of a full-body PET based on liquid xenon, read out by silicon photomultipliers. An end-to-end \textsc{Geant4}-based Monte Carlo simulation is used, including the optical transport of scintillation photons, to assess the performance of the reconstruction of the gamma interactions in the medium.

The intrinsic performance of LXe scintillation, together with SiPM read-out, is excellent: a CTR of $\sim$60 ps FWHM is obtained for point-like gamma interactions, while the rest of interactions shows a CTR of around 370 ps FWHM.  The time jitter values due to the electronics and the sensor response are estimated to be 30 and 40 ps $\sigma$, which results in a combined fluctuation of $\sim$ 117 ps FWHM. Adding the smearing due to these instrumental effects, a global CTR of 350 ps FWHM is obtained. These results indicate that, if improvements are made in the SiPM and ASIC technologies, using techniques to select \textit{photoelectric-like} events, such as, for instance, DNN-based algorithms, could further improve the CTR.


The resolution in the determination of the gamma interaction points in LXe is around 1 mm for point-like gamma interactions and between 1 and 2 mm for the whole ensemble of events, with large tails. This very good performance results in an excellent resolution of the image of a point source: a spread of around 1.2-1.3 mm FWHM in the three coordinates is found, dominated by the gamma non-collinearity.


\acknowledgments
This work was supported by the European Research Council under grant ID 757829 and by Ministerio de Econom\'ia y Competitividad for grant FPA2016-78595-C3-1-R.


\end{document}